\documentclass[aps,prl,twocolumn,groupedaddress,showpacs,floatfix]{revtex4}
\usepackage{graphics}
\usepackage{epsfig}
\begin{document}
\title{Total energy calculation of high pressure selenium: The origin of 
incommensurate modulations in Se-IV and the instability of proposed Se-II}
\author{G.J.Ackland and H.Fox}
\affiliation{School of Physics and Centre for Science at Extreme
Conditions, The University of Edinburgh, Mayfield Road, Edinburgh,
EH9 3JZ, UK.}

\begin{abstract}

We present calculation of the high pressure crystal structures in
selenium, including rational approximants to the recently reported
incommensurate phases.  We show how the incommensurate phases can
be intuitively explained in terms of imaginary phonon frequencies
arising from Kohn anomalies in the putative undistorted phase.  We
also find inconsistencies between the calculated and experimental
Se-II phase - the calculations show it to be a metastable metal
while the experiment finds a stable semiconductor.  We propose that the experimentally reported structure is probably in error.
\end{abstract}
\date{\today}
\pacs{61.50.Ks,62.50.+p} \maketitle

The group VI elements selenium and tellurium are semiconductors
with a hexagonal crystalline structure at ambient pressure.  Both
elements undergo a series of phase transitions as pressure 
is increased, transforming to various complex structures
denoted by roman numerals I, II, III, IV. The generally accepted
crystal structures associated with these phases in selenium is
hexagonal Se-I; monoclinic Se-II from 14GPa; triclinic Se-III
from 23GPa;  orthorhombic Se-IV from 28GPa;
and subsequently simple higher symmetry phases\cite{akahama2}.
Tellurium undergoes a similar set of phase
transitions at lower pressures, without the equivalent of
Se-II.

In 2003 new work using combined powder and single-crystal x-ray
diffraction\cite{henjy} found that the Te-III phase had a more complex
structure than had previously been thought, involving a modulated
distortion of the unit cell.  It was suggested that Se-IV had a similar
structure experimental confirmation of this is
in progress\cite{mimetal}.

Density functional theory\cite{HKS,KS}, using pseudopotentials, plane
waves and local exchange correlation functionals has been the method
of choice for theoretical study of selenium.  Within the local density
approximation(LDA) the $p$-bonded covalent chain structure and
anomalous compression mechanism (one lattice parameter increases with
pressure), was demonstrated\cite{hadi}. The anisotropy showed that the
LDA introduced an error equivalent to a pressure, which can be
improved by including gradient corrections (GGA)\cite{Kresse}. Since
then band structures of the complex high pressure phases have been
investigated using the experimental parameters without full relaxation
of the crystal structure or incommensurate modulation\cite{geshi}, and
the liquid phase has been investigated up to high
pressures\cite{liqSe}.

Although there are several papers on individual crystal structures,
proposed transition mechanisms, and the simple high symmetry
structures at pressures above Se-IV\cite{rudin}, we are not aware of
published calculated phase transition pressures in
SeI-IV\cite{except16}. This may be because of their structural
complexity, because the enthalpy differences between the phases are
extremely small, or, because of imperfect agreement with the reported
sequence of transitions.

In this letter we report density functional calculations of the
first four phases of selenium, identifying the stable structures
over the pressure range to 40GPa. Subsequently we investigate
the nature of the incommensurately-modulated fourth phase of Se,
approximating the incommensurate structure and showing it to be an
extreme case of a Kohn anomaly.


We used the computer program VASP\cite{vasp}, treating exchange
correlation effect with a GGA
which is essential to obtaining accurate pressures for the low density
structure\cite{hadi,Kresse}. The wavefunctions are sampled using the
method of Monkhorst and Pack, and the energies converged to 1meV with
respect to k-point density and 0.2meV with respect to plane wave
cutoff.  For Se I, II and IV a k-point mesh of 15$^3$ was sufficient
while for Se-III we used 17$^3$.  Each structure was fully relaxed
with respect to lattice parameters and internal coordinates according
to \emph{ab initio} calculated forces and stresses.  The core region
is described with ultra-soft psuedopotentials\cite{vanderbilt}.  For
consistency, we performed calculations on reported structures of all
phases I through IV.
To calculate band structures a self-consistent calculation using
the aforementioned meshes was first carried out, to obtain the ground
state charge density, and then the single particle states of the
Kohn-Sham Hamiltonian were evaluated non-self consistently on a finer mesh..



Se-I is the ambient pressure phase of selenium. It has a hexagonal
unit cell, with the atoms arranged in helical chains running
parallel to the \emph{b}-axis. 
Each unit cell contains three atoms, in positions    $( 0, 0, u )$ $( u,\frac{1}{3}, 0 )$ $( -u, \frac{2}{3} -u )$ with u=0.289 at 20GPa.
The equilibrium lattice parameters at this pressure are a=c=3.569\AA; b=5.135\AA.
In accordance with previous work\cite{hadi,Kresse,sjc}, we find that as the
pressure is increased, the size of the \emph{b} parameter at first
increases, then starts to contract at 10GPa. However this change
is a tenth smaller than the size of the reduction in the other two
directions. 


Se-II is reported to have a monoclinic structure, with six atoms
per unit cell, at positions $(0,0,0)$, $(u,0,v)$,  $(-u,0,-v )$
and at ($\frac{1}{2},\frac{1}{2},0$) relative to these. 
At 22GPa, u=0.147 and v=0.339. The equilibrium
lattice parameters of the cell at this pressure are
  \emph{a}=6.562\AA  \emph{b} = 2.665\AA  \emph{c}=6.326\AA  $\beta$=105.3$^o$
This is a layered structure, with each atom closely bonded to four
other atoms. On the increase of pressure, the parameter \emph{a}
reduces much faster than \emph{b} and \emph{c}  as the 
layers are being pushed together. The angle $\beta$
changes very little over the calculated pressure range.

\begin{figure}
\includegraphics[width=\columnwidth]{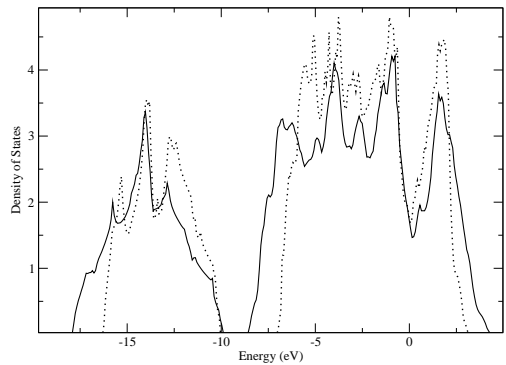}
 \protect{\caption{Calculated density of states for Se II structure
from a $25^3$ k-point grid at 0GPa (dotted line) and 22GPa (solid
line) with ions and unit cell fully relaxed within the spacegroup
reported experimentally.  The Fermi energy has been shifted to zero,
although it lies at a minimum in the density of states, no band gap is present
and this structure would result in a metallic material\cite{nogap}.  
\label{SeII_bs}}}
\end{figure}

Fig.\ref{SeII_bs} shows the density of electron states of Se-II at 2GPa and
22GPa.  The effect of pressure is primarily to broaden the distribution. Consistent with previous work, there are two sets of bands,
corresponding to 4p electrons\cite{16}. There is no band gap,
showing that at either pressure the material is a metal\cite{nogap}.


Se-III is reported to be triclinic, with six atoms per unit cell. 
At 28GPa, the equilibrium lattice parameters of the cell are
  =3.460 \AA      b=12.240 \AA         c=2.624 \AA
$\alpha$=91.085$^o$     $\beta$=113.445$^o$        $\gamma$=89.608$^o$.
Se-III also has very anisotropic compression, the largest decrease in
the cell is in the \emph{a} direction, which decreases by 10\%
over a range of 20GPa, compared with 5\% for c and only 1.5\% for
b. Angles $\alpha$, $\beta$, $\gamma$ change little as the
pressure is increased. At 24GPa, the calculated atomic positions  are:   
{(-0.003,-0.001, -0.024)} {(0.465. 0.167. 0.389)}   {(0.030, 0.331, 0.064)}
and {( $\frac{1}{2},\frac{1}{2},\frac{1}{2}$)} relative to these.
Band structure calculations show that Se-III is metallic.

\begin{figure}
  \includegraphics[width=\columnwidth]{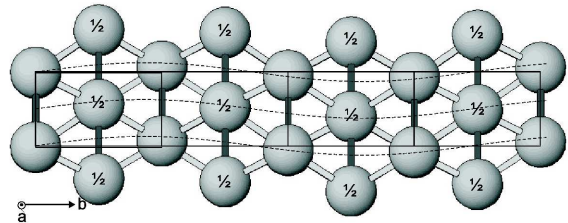}\\
  \caption{Four cells of the modulated body-centered monoclinic cell
  of Se-IV viewed down the \emph{a} axis.  The dashed line shows the
  modulation at the experimental value of \emph{q}=0.28, the
  calculations were done on cells with this modulation adjusted to
  $q=\frac{1}{3}$ and $q=\frac{1}{4}$ corresponding to a modulation
  wavelength of 3 and 4 cell respectively.  Atoms labelled with
  $\frac{1}{2}$ lie at that fractional coordinate in the c direction,
  others lie in the plane. (Figure courtesy of
  M.I.McMahon)}\label{pic_seIVb}
\end{figure}

We modeled the Se-IV cell by relaxing ions and lattice parameters from 
the primitive cell reported experimentally\cite{henjy}
without the incommensurate distortion (Fig.\ref{pic_seIVb}). At
35GPa, the equilibrium lattice parameters were a=3.383\AA,
b=4.071\AA,  c=2.613\AA and     $\beta=115.644^o$
This compares with the experimental results:
    a=3.299\AA,  b=3.996\AA,
    c=2.587\AA, and    $\beta=113.105^o$.
The calculated unit cell parameters are about 2\% larger than the
experimental value at the same pressure.  This is typical for the
GGA.





It is impossible to do calculations on an incommensurate cell
using periodic boundary conditions, however one can use a rational
approximant\cite{skr} assuming that a distortion close to the
observed $q$ will also be stable relative the undistorted
structure. Here we approximated the distorted
structure with tripled, quadrupled or heptupled cells corresponding to
q=$\frac{1}{3}$, $\frac{1}{4}$ and $\frac{2}{7}$.  In each case the calculation
was started with frozen-in values of the incommensurate wave
displacement vector taking  Te-III data $u_x=0.0249$ and $u_z=0.1026$ as
initial symmetry-breaking displacements.

For q=$\frac{1}{4}$ calculations  were done at four different
pressures: 25, 30, 35, and 40GPa. In all cases, the displacement
remained and, the enthalpy of each atom in the distorted structure
is 3meV smaller, with
the distorted structure being slightly larger than the undistorted
one. Table.\ref{q4} shows a summary of results found. The energy
differences to the undistorted phase are small, but lie within k-point 
and cutoff convergence.
We tested this to greater accuracy  which can be obtained by
eliminating sampling errors for the crucial energy
\emph{differences} by comparing cells with and without the
distortion, using identical k-point and cutoffs\cite{aak}.

\begin{table}
  \centering    \begin{tabular}{|l||ll|ll|l||ll|l|}
\hline
    P &   $q=\frac{1}{4}$ & &  &  & & $q=\frac{2}{7}$ &  &\\
    (GPa) & $u_x$ &$v_x$ & $u_z$ &$v_z$ & $\Delta H$ & $u_x$ &$u_z$ & $\Delta H$\\
\hline
    25 & 0.017 & 0.012& 0.057 & 0.040 & -3 &
     0.021 & 0.077 & -4 \\
    30 & 0.016 & 0.011 & 0.057 & 0.040 & -3 &
     0.018 &  0.070 & -3  \\
    35 &  0.014 & 0.010 & 0.057 & 0.040 & -3 &
      0.015  & 0.065 & -3 \\
    40 & 0.012 & 0.008& 0.055 & 0.038 & -2 &
     0.012 & 0.060 & -2 \\
\hline  \end{tabular}
  \caption{Details of frozen phonons at q=$\frac{1}{4}$ and
q=$\frac{2}{7}$ at various pressures: amplitudes of modulations in the
xz plane for q=$\frac{1}{4}$ correspond to maximum displacement $u$,
intermediate displacement $v$ and two atoms fixed at unmodulated
positions. For  q=$\frac{1}{4}$ we have chosen the frozen
phonon to be in phase with the unmodulated cell: within error the
ratio $u/v=\sqrt{2}$ implies that the distortion can be described as a
single harmonic phonon.  For q=$\frac{2}{7}$ the displacement pattern
is again harmonic, $u_x$ and $u_z$ represent the amplitude of a fitted
sine wave, but the phase of the modulation is not locked.  
Enthalpy differences are of order 2-3 meV/atom (10-30meV/cell) 
from the unmodulated SeIV calculated with the same size, kpoint sampling 
and supercell.  The close match between modulated and unmodulated cells 
means that the enthalpy differences are significant. 
}\label{q4}
\end{table}

The calculated distortions at $q=\frac{1}{4}$  and  $q=\frac{2}{7}$ 
are smaller than those found experimentally at 35GPa.  This is to be
expected since the calculated distortions are not
the most favored one.  Likewise, the very small enthalpy
difference (-3meV/atom) is a lower bound on the stability of the
incommensurate wave.  The displacement eigenvector 
(transverse, $u_z/v_z=0.22\pm0.04$ at 40GPa is in excellent agreement 
with the experimental value (0.26 at 42GPa\cite{mimetal}).

For q=$\frac{1}{3}$ we carried out calculations at six different
pressures: 10, 20, 25, 30, 35, and 40GPa, the last three when
Se-IV was predicted earlier to be stable. At all six pressures,
the structure relaxed back to the unperturbed structure, as did 
calculations with $q=\frac{1}{7}$.

\begin{figure}
\protect{\includegraphics[width=\columnwidth]{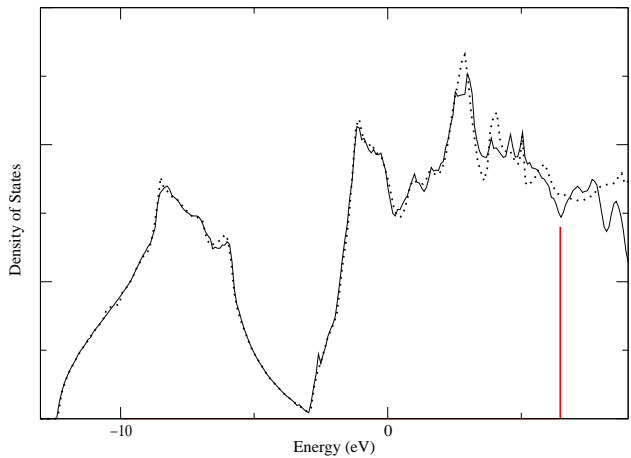}}
\protect{\caption{Calculated Kohn-Sham band structure and density
of states for Se IV at 35GPa with (solid line) and without (dotted line) 
the modulation at $q=\frac{1}{4}$. 
The Fermi energy,  $E_F=6.46eV$ in both cases, is shown by the vertical line. 
The Kohn anomaly arises because the distortion perturbs states at 
$(0,\frac{1}{4},0)$ which lie close to $E_F$, leading to 
the drop in the density of states at $E_F$ \label{BS_dosIV}}}
\end{figure}

While DFT is clearly a sufficient theory to explain the incommensurate 
modulation, a more intuitive explanation exists:
modulation arises from a
Kohn anomaly (interaction between nested Fermi surface and a
phonon) so large that the phonon energy becomes negative and the soft mode ``freezes in''. 
Experimentally\cite{henjy}, the
incommensurate q-vector drops from 0.31-0.28 between 30 and 50GPa,
remaining at 0.28 until 70GPa\cite{mimetal}.
A standard
Kohn anomaly in the free electron picture would occur at $q=2k_F$,
and soften (i.e. reduce the frequency of) the phonon at that
wavevector. Here the softening of the mode is so great that it
freezes in (i.e. the frequency is imaginary), lowering the
symmetry of the unit cell and allowing coupling to the unit cell
parameters.  The stability arises because the phonon mode opens a 
pseudogap at the Fermi energy, which can been seen as a dip in the 
electronic density of states:
Fig.\ref{BS_dosIV} shows that the distortion has exactly this effect. 
We examined the band structure of the undistorted cell
and find that the calculated Fermi surface cuts the band in the 
$(0,k,0)$ direction at values of $k$ rising from 
0.1 at 20GPa to 0.26 at 40GPa, while the distorted structure 
has a pseudogap along this direction.  
Although this does not give us the exact value of $q$
from the $2k_F$ of the undistorted phase, the agreement is
sufficiently close to be confident of the mechanism.

\begin{figure}
  \includegraphics[width=\columnwidth]{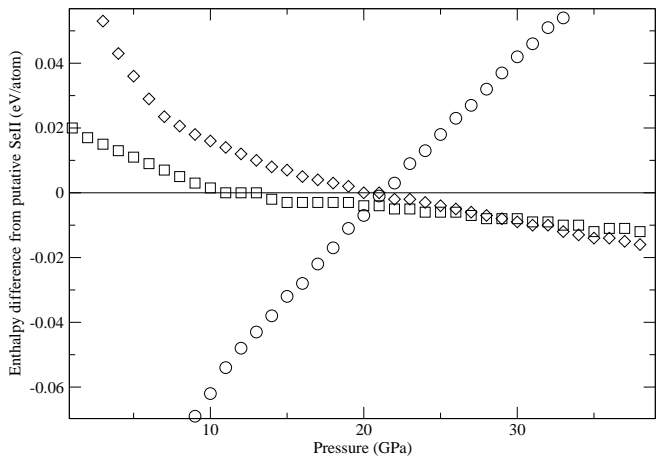}
  \caption{Calculated enthalpy differences between Se II and other
  phases.  The stable structure at is the one with the lowest enthalpy
  at each pressure. Circles denote Se I, squares Se III and diamonds
  unmodulated Se IV.  Errors are of order the size of the
  symbols}\label{se_en}
\end{figure}

Fig.\ref{se_en} shows the phase stability deduced from our calculations of
enthalpy of each phase from 1 to 40GPa. The stable
structure of selenium at each pressure is that with the lowest
enthalpy, and although the differences in
enthalpy are small throughout the whole pressure range they are above 
our convergence errors.

The most notable feature is that our structure for 
Se-II, reported experimentally at 14GPa\cite{akahama1,bundy,ohmasa},  
is not predicted to be
stable at any pressure. It is higher in energy by around 8meV/atom than
Se-III, which is greater than convergence of the
calculation. This may be because of an entropic effect: a symmetry
breaking distortion may have lower energy with Se-II a dynamically
stabilized phase with a transition below room
temperature\cite{rev,ndd} (8meV corresponds to a temperature of
100K).  However, this would not explain the band structure
observed in the calculation - along with previous
authors\cite{geshi,ikawa} we predict that Se II is metallic while
experimentally it is a semiconductor. This is a strong indication
that the atomic positions used in the calculation (starting from
the experimental ones) are incorrect and we conclude that there may
be an error in the experimentally reported crystal structure

We predict a phase transition from hexagonal Se-I direct to
monoclinic Se-III at 21.5GPa accompanied by a volume 
reduction of 5.5\%;  experimentally Se-III is
observed at 23GPa\cite{bundy,akahama2}.
The transition from Se-III to Se-IV is predicted at 27.5GPa,
compared to the experimental value of
28GPa\cite{akahama1,parth,ohmasa}. These excellent agreements give us confidence that our method correctly describes high pressure selenium.


To  summarize, {\it ab initio} total energy and band structure 
calculations on four different phases of selenium are
reported. The generally accepted sequence of phase transitions is
not reproduced although unit cell and internal parameters  are
in good agreement with experiment.  The calculated instability of the
experimentally reported Se-II structure, combined with our
prediction that it should be metallic, leads us to suspect that
this phase may have been mischaracterized, and there is very 
recent evidence for unexplained diffraction peaks in Se-II\cite{MIM}.

Body-centered monoclinic Se-IV is unstable against a phonon
modulation with $q=\frac{1}{4}$ or  $q=\frac{2}{7}$. Inspection 
of the band structure of the
undistorted phase leads us to conclude that this instability arises
from a Kohn anomaly in the phonon spectrum of sufficiently large
magnitude that the phonon eigenenergy is negative (i.e. the mode
is unstable).  The wavevector of the nested Fermi surface is
incommensurate with the crystal structure and varies with
pressure. This gives a theoretical picture of the recent
unexplained experimental observation of an incommensurate Se-IV.

\section{Acknowledgements}
We thank M.I.McMahon for many useful discussions and making 
data available to us prior to publication, and EPSRC for computer resources.


\begin{thebibliography}{100}
\bibitem{nogap} DFT is known
to underestimate bandgaps and it is conceivable that this could cause
a spurious bandgap closure.  However, this would have the effect of
overstabilizing the Se II phase by {\it lowering} the enthalpy, the
opposite of what is observed in Fig\ref{se_en}.
\end{thebibliography}
\end{document}